# Autonomous Vehicles for Smart and Sustainable Cities: An In-Depth Exploration of Privacy and Cybersecurity Implications

By

**Hazel Si Min Lim and Araz Taeihagh\***

\* Correspondence: Araz Taeihagh, Lee Kuan Yew School of Public Policy, National University of Singapore, 469B Bukit Timah Road, Li Ka Shing Building, Singapore 259771, Singapore spparaz@nus.edu.sg; Tel: +65-6601-5254

# Autonomous Vehicles for Smart and Sustainable Cities: An In-Depth Exploration of Privacy and Cybersecurity Implications

**Hazel Si Min Lim and Araz Taeihagh\***

* Correspondence: Araz Taeihagh, Lee Kuan Yew School of Public Policy, National University of Singapore, 469B Bukit Timah Road, Li Ka Shing Building, Singapore 259771, Singapore spparaz@nus.edu.sg; Tel: +65-6601-5254

**Abstract:** Amidst rapid urban development, sustainable transportation solutions are required to meet the increasing demands for mobility whilst mitigating the potentially negative social, economic, and environmental impacts. This study analyses autonomous vehicles (AVs) as a potential transportation solution for smart and sustainable development. We identified privacy and cybersecurity risks of AVs as crucial to the development of smart and sustainable cities and examined the steps taken by governments around the world to address these risks. We highlight the literature that supports why AVs are essential for smart and sustainable development. We then identify the aspects of privacy and cybersecurity in AVs that are important for smart and sustainable development. Lastly, we review the efforts taken by federal governments in the US, the UK, China, Australia, Japan, Singapore, South Korea, Germany, France, and the EU, and by US state governments to address AV-related privacy and cybersecurity risks in-depth. Overall, the actions taken by governments to address privacy risks are mainly in the form of regulations or voluntary guidelines. To address cybersecurity risks, governments have mostly resorted to regulations that are not specific to AVs and are conducting research and fostering research collaborations with the private sector.

## 1. Introduction

To make cities inclusive, safe, resilient, and sustainable is one of the goals set out by the United Nations (UN) in their 2030 Sustainable Development Goals (SDGs). Cities are now home to over half the world's population, and by 2050 it is expected that cities will account for two-thirds of the world's population [1]. As cities grow rapidly, so does the demand for mobility and its attendant pressures on the economy, social stability, and the environment. Traffic congestion cost the US economy $124 billion in 2013, projected to increase to $186 billion by 2030 [2]. Other growing problems that the transportation industry contributes to are air pollution, climate change, negative effects on public health, and unequal access to services [3]. 'Smart' and 'sustainable' cities can help to alleviate these problems by utilising technological innovation to help individuals, businesses, and governments attain higher living standards while ensuring the sustainability of social and environmental systems [4]. Sustainable transport systems and infrastructure can reduce the negative impacts of urban development on the environment, economy, and society [5].

Autonomous vehicles (AVs) have emerged as a potential solution to modern day transport problems. Widespread AV adoption can reduce environmental degradation through reduced emissions and energy consumption while providing beneficial economic and social outcomes through improved efficiency, traffic flow, road safety, and accessibility to transport, among other benefits [6–9]. Much of these benefits stem from AVs' connected nature, which enables them to communicate with other vehicles and critical infrastructure to optimise traffic and maximise all associated benefits for sustainable and smart cities [10,11].

Since the introduction of AVs in 2010, their development and appeal has increased significantly. However, the successful operation of AVs and their impact on society depend significantly on their management and on addressing risks associated with them. Two of these risks are privacy and cybersecurity. Firstly, the ability of AVs to store and communicate personal data may conflict with data privacy laws, creating ambiguity regarding the ownership, storage, and transmission of data collected in the AV [12]. Secondly, AVs face major cybersecurity risks if the communication networks, which are crucial for their safe operation, are not secure from hacking. Unauthorised access to these networks can have dire consequences, such as undermining a vehicle's safety and utilising personal data for malicious intent [7]. With increasing emphasis on "safe, secure, and affordable access to global networks" as a key criterion for social and economic progress, ensuring cybersecurity and privacy while "expanding Internet access and instilling human rights" is vital for sustainable development [13]. Other risks include safety risks, liability risks, and employment risks [14–16].

In this study, we focus on privacy and cybersecurity risks of AVs, which can potentially hinder the progress towards greater sustainability if they are not addressed appropriately by governments. Scholars have provided recommendations to address AV-related privacy and cybersecurity risks [17–21], but only a limited number of studies have explored the link between privacy and cybersecurity risks of AVs, and smart and sustainable development. Furthermore, there is a growing but limited literature that provides country-specific analysis regarding the governance responses to AVs. This article aims to address the following questions: (a) Why are AVs important for sustainability? (b) What are the privacy and cybersecurity risks associated with AVs? (c) What are the implications of these risks for sustainability? (d) What are the emerging solutions to address these risks?

The next section provides an overview of the literature on smart and sustainable cities, as well as background information about AVs. In Section 3, we discuss why AVs are important to achieve smart and sustainable development. We then explore the implications of privacy and cybersecurity of AVs on smart and sustainable cities and examine the governance responses to manage them in Sections 4 and 5, before concluding.

## 2. Background Information

### 2.1. Sustainable and Smart Cities

The idea of a 'sustainable city' first originated as a reference to "development that meets the needs of the present without compromising the ability of future generations to meet their own needs" [22]. A city is said to be sustainable if it can realise a higher standard of living while retaining "a self-sufficient economic, social, and environmental system" [4]. Sustainability is not a goal, but "a process of continuous improvement according to the needs and the context, which can vary in time and space" [5]. The term gained popularity in the 1990s and is widely understood by reference to the strong interrelationship between economic growth, environmental protection, and social equity, also known as the 'triple bottom line' [4,23]. While some scholars place greater emphasis on environmental aspects, others view sustainability from a socio-economic angle by emphasising the achievement of social equity [24,25]. However, advocates of the 'triple bottom line' often frame sustainability in broader terms, such that environmental considerations are in less conflict with social and economic sustainability [4].

The importance of building sustainable cities has received global attention as reflected in the UN's 2030 SDGs, which aims to promote inclusive and sustainable economic growth, industrialisation, and innovation, to ensure good health and wellbeing for all ages, and to provide affordable and clean energy to meet business demands while protecting the environment [1]. More recently, the term 'smart city' has emerged as a new concept amidst the acceleration of competition among cities for businesses and talent, which has driven cities to place more emphasis on "economic feasibility and engineering systems solutions" to realise their desired objectives of urban development [4]. The first conception of a smart city emphasised the need to implement new and innovative technologies to solve problems in urban areas, and later emphasised a "greater degree of involvement of the local authorities" regarding the application of these technologies [26]. These smart technologies are implemented in the cities' infrastructure, as well as in the cities' governance and education systems, to improve interaction between citizens and their government [23]. In the latest concept of smart cities, Smart City 3.0, scholars advocate for citizens' active usage of smart solutions to improve living standards and sustainability [23,26]. Caragliu et al. [27] identified six main characteristics of smart cities: utilising networked infrastructure for socio-economic development; adopting business-led development; ensuring equitable growth; capitalising on human talent and creative industries; the need for social capital for adaptation and innovation; and environmental sustainability.

Due to its emphasis on connectivity as the main source of growth, the 'smart city' tends to shift attention away from environmental considerations and more towards infrastructure and information use; bibliographic analysis of the literature has found that the use of the term 'smart city' has exponentially increased since 2009 and in 2012 even surpassed that of 'sustainable city' in academic literature, suggesting that this new concept is likely to be the main driver of urban sustainability [4].

*2.2. Background to AVs*

AVs (also known as driverless vehicles) are able to make decisions independently of human interference, in the face of uncertainty, and are set to revolutionise the transport industry [16,28] (we use the terms AV and driverless vehicles interchangeably in this article). AVs rely on sensor data and artificial intelligence (AI) to interpret the data, to make decisions regarding vehicle operation, and to adapt to changing conditions [29,30]. The advantage of AVs stems from their ability to rapidly process information and to adapt to their environment much faster than a human, and exchange information through vehicle-to-vehicle (V2V) and vehicle-to-infrastructure (V2I) communication technologies [29].

The independence from human judgement and superior awareness of road conditions make AVs appealing in terms of safety, as they are mostly associated with the elimination of human error that has been responsible for countless fatalities in the history of motor vehicle accidents. Over 35,000 fatalities occur from traffic accidents in the US every year [31]; China experiences around 260,000 road fatalities a year; Japan experiences around 4000; and globally there are over a million road fatalities every year [29,32]. Adopting driverless cars can potentially reduce, if not eliminate, the most significant cause of car accidents (human errors) [14]. Studies estimate that a significant number of crashes could be avoided as a result of autonomous braking [33]. AVs are also capable of avoiding collisions while complying with traffic rules if the latter are made precise and unambiguous [34,35].

Given the rapid developments in AV technology, established car manufacturers and software companies are competing to harness new opportunities in the car- and ride-sharing industry [30,36,37]. A challenge faced by most governments is balancing the desire to adopt AVs and addressing the risks they introduce. AVs can potentially introduce great societal and economic benefits through increasing safety and mobility, particularly for the disabled and elderly. In addition, it is claimed that they can potentially increase competitiveness, productivity, and fuel economy, as well as reducing congestion and pollution [38–40].

However, AVs involve different types of risks (for a more comprehensive study of the societal implications of AVs see [41]). In this paper, the focus is on privacy and cybersecurity risks introduced by AVs. Despite the elimination of driver error, risks may arise from a myriad of factors, such as system errors, cyber-attacks on safety systems, and less cautious behaviour of both passengers and pedestrians. Sophisticated data processing and storage abilities of AVs raise data privacy concerns. Connection to external networks, which is necessary for vehicle cooperation on roads, exacerbates privacy risks as data can be easily retrieved once a network is compromised. In addition, the responsibility for car accidents will shift from the occupants to manufacturers, as the autonomous system, not the human, is in control of the AV. Both manufacturers and software providers face greater risk of lawsuits arising from accident compensation, which may deter innovation if liability laws fail to systematically apportion responsibility.

The Society of Automotive Engineers (SAE) defines five levels of vehicle automation: at levels 1 and 2 (assisted automation and partial automation, respectively), the human driver performs the driving operations [42], while at levels 3 to 5, the automated driving system carries out all of the dynamic driving tasks. It is expected that the human driver can control the vehicle occasionally at level 3 (conditional automation). The distinction between levels 4 (high automation) and 5 (full automation) is that only at level 5, the vehicle is expected to be able to operate under all environmental conditions [41]. In our study, we use the SAE classification, which is the most widely adopted classification of AVs, and focus on fully autonomous AVs (SAE levels 4 to 5), as they entail a greater shift in the society.

## 3. AVs and Sustainability

AVs offer potentially transformative benefits that contribute to economic and social sustainability. Firstly, AVs can increase economic efficiency by increasing road capacity, improving traffic flow, and reducing congestion [41,43–46]. These benefits stem mostly from AVs' connectivity to external communication networks, known as V2V and V2I communication networks [28], which enable them to manage and distribute data on the go [30,47]. This enables platooning, which increases spatial efficiency, road capacity, and congestion [48]. Simulation studies have shown that the positive effects on congestion increase as the platoon length and cooperation between AVs increase [41,49]. The economic benefits are significant, considering the high costs of congestion, which amount to over $120 billion and £30 billion a year in the US and UK, respectively [29,50]. On the other hand, it is unclear when and whether these benefits will emerge, as the opposite may occur if AVs increase overall traffic demand, if AV penetration rates are low, or if road capacity does not increase as expected [51,52]. In addition, AVs may lead to a decrease in road accidents due to the absence of human error, which will also improve traffic flow [48]. Congestion can be further improved with upgrades to existing infrastructure to complement AVs, such as intelligent traffic lights that respond to instantaneous demand on roads and developing automated crossings that do not need traffic lights to operate [30,46]. Due to AVs' connected nature, the locations of repair slots are easily communicated to AVs to optimise and centralise maintenance [30], yielding time savings.

AVs can provide ride-sharing services to yield greater economic and social benefits. The 'sharing economy' [48,53,54] enables the reallocation of underutilised resources for more productive uses, unlocking new sources of supply at a lower cost [53]. With AV technology, manufacturers may provide driverless car services by selling AV hours rather than the AV itself [53], also known as 'shared' AVs [18]. It is suggested that shared AVs are likely to be more efficient than providing ride-sharing services, such as Uber, Grab and Lyft [18,48]. Hörl et al. [30] note that fleet operators can buy AV hours optimised for "dense downtown situations or rural areas", tailoring transport solutions that are unique to the community. In addition, using AVs frees up time, allowing passengers to complete work in the vehicle online, rather than at a centralised office all the time [30]. AVs can also significantly increase productivity in labour-intensive transportation industries and related occupations by reducing labour costs, although employment concerns may arise in the short-term, e.g., the freight industry [30], food distribution industry [55], and traffic policing [56].

Secondly, AVs can allow for more inclusive economic growth through greater accessibility to transport. AVs open up accessibility to new user groups who were previously unable to drive, such as the elderly and handicapped, which makes it possible to achieve a more inclusive society [30,57,58]. Furthermore, shared AV services can solve the first- and last-mile issue and increase spatial availability as they are able to drive to users autonomously [30,59]. It must be noted, however, that the registration of such shared mobility services relies on existing user access to digital platforms, which could be an additional barrier to mobility for those disadvantaged by the "digital gap or lack of access to banking" [3]. Greater accessibility may also worsen congestion through induced demand [48,59]. In addition, shared AVs may operate under a more profit-driven business model that seeks to develop its customer base while side-lining other more sustainable but less profitable means of transport, such as walking or cycling [3]. In the case of individual ownership, AVs could be programmed to "cruise around the block while waiting for the owner to finish their business", which would contribute to congestion [53]. It is suggested that this problem can be solved if AVs join other transportation services, such as taxis, ride-sharing pool systems, and small buses, to create "local links" in the transportation network; However, AVs risk cluttering up the areas around transit stations, where "walkability is the core feature sought by local governments, business investors, and communities" [59]. Hence, there is also a need to balance the benefits of greater accessibility with the negative aggregate effects arising from more unsustainable individual practices that shared mobility services may encourage [3].

Thirdly, AVs can have a potentially beneficial impact on public health and wellbeing. van Schalkwyk and Mindell [60] summarised the impacts of different transport modes on health. For instance, mental health is significantly affected by the lack of accessibility to transport modes and noise pollution stemming from excessive car usage, and poor transport planning may have a detrimental effect on the safety of residents, exacerbate urban sprawl, and disadvantage those with disabilities, whereas transport modes that encourage greater physical activity can reduce the risks of stress, depression, and various diseases [56,60]. AVs, by increasing accessibility, independence, and road safety, can have a beneficial impact on public health, especially for the elderly as AVs can help the elderly overcome physiological barriers to mobility and reduce the risk of collision for ageing drivers [56]. In addition, AVs can potentially reduce non-communicable diseases by reducing driving-induced stress, which is "a key contributor to hypertension" [56]. However, it is unclear whether these benefits will be offset by some unintended consequences AVs may have on public health. For instance, AVs may lead to continued decreases in physical activity, which can result in increases in obesity and consequently other diseases if individuals increase the usage of AVs at the expense of active modes of transport or other physical activity throughout the day [56], worsening the effects of a sedentary lifestyle. Public authorities may capitalise on the connectivity of AVs to increase public finances and improve social welfare, such as monitoring travel to impose taxes and to construct a socially-optimal traffic environment [30,61].

In addition to the economic and social benefits, AVs can potentially reduce the negative environmental impact of the transportation industry and thereby contribute to environmental sustainability. Studies have shown that AVs can reduce carbon emissions and improve fuel economy [41,62–64]. In addition, AVs' ability to communicate and coordinate with other vehicles reduces traffic congestion and idling, thereby reducing unnecessary acceleration and braking [41,65]. Platooning also improves aerodynamics, reducing fuel consumption and emissions [19,66,67]. Currently, it is uncertain whether these positive effects on the environment will be offset by other unintended consequences. The US Department of Energy [68] has estimated that fuel usage and related emissions could either decrease by 60% or increase three times [69]. The latter may be the case if the induced vehicle demand from AVs contributes to higher emissions levels [67,70,71]. In addition, improved fuel efficiency depends on other factors, such as the design of control algorithms that account for fuel efficiency, and the penetration rate of AVs [72,73]. Milakis et al.'s [41] literature review on the impacts of AVs suggests that in the short-term AVs can improve fuel savings and reduce emissions, but their long-term effects are ambiguous. Li et al. [64] highlight that emissions and energy consumption may be reduced only if V2I technologies are applied correctly to manage traffic flow at intersections. Others, such as Newman et al. [59], argue for a more sceptical perspective regarding AVs' overall impact on sustainability and show that AVs still cannot compete with individual mobility due to the greater independence/freedom of the latter, and cannot compete with transit-based mobility, due to higher vehicle occupancy/capacity of the latter.

## 4. AV Privacy

### 4.1. AV Privacy Implications for Smart and Sustainable Cities

AVs are one of many smart devices that can store highly sensitive data through video and audio recording [21,74], and that can also transmit such data to other vehicles, connected infrastructure, and third-party organisations through external V2V and V2I communication networks. While it is recognised that such data sharing is essential for efficient traffic management [12] and the accurate assignment of liability in the event of collisions [75], unrestricted sharing of data raises privacy concerns.

Informational privacy of AVs is important in order to reap the full benefits of increased connectivity in an information society [76], and thus to develop smarter and more sustainable cities. Informational privacy (or data privacy) is defined as "the protection of a person and his/her behaviour" such that the individual is "able to control the risks for his or her rights to privacy, freedom, or equality caused by the processing of data related to him or her" [77]. Currently, it is unclear who can access and use the data collected in AVs [28]. Firstly, informational privacy in AVs is important to safeguard against the misuse of personal information that can inflate societal discrimination, which is crucial for sustainability from the perspective of social stability. As there are "no explicit rules to consider certain data special and have special hindrances for their usage", the data collected in AVs may be misused in many ways that disadvantage AV passengers [78]. For instance, insurance companies and credit rating agencies can use AV data to calculate insurance premiums and credit scores associated with individuals, which have been shown to be very inaccurate [78]. In general, highly sensitive personal information, such as geographical locations, may be used to make unfair inferences about individuals in systems that use AI such as AVs, especially if existing datasets are biased against people of certain ethnicities or sexualities, and consequently exacerbate existing inequalities in society [79,80]. Scholars have highlighted other examples of data misuse that could be detrimental to AV users. For instance, past travel in AVs can be used to predict the behaviour of AV users and to harass them through tailored advertising and marketing strategies [12,17]. An important implication is the increasing power disparity between the organisations that control such data and individuals in society [17]. In particular, humans leave behind unique signatures in data relating to their geographical location, which can be used for re-identification with the aid of a small amount of side-information (side information is defined as the information an adversary acquires from indirect sources such as social networks, news articles, and chat logs to infer a target's location through a set of anonymous traces [81]) [12,81,82]. Nevertheless, Fagnant and Kockelman [83] highlight the importance of balancing privacy concerns against the considerable social and economic benefits from data sharing, such as improving transport planning and investment decisions.

Furthermore, informational privacy in AVs is crucial to building consumer trust, which is crucial for economic sustainability. This is especially so in the sharing economy, where consumers act as "prosumers" who are "offering mobility services themselves" [54,77]. Only where there is consumer trust can there be sufficient market uptake of AVs, which is crucial if further improvements are to be made to the technology in terms of safety, efficiency, traffic flow, and other benefits to achieve a smarter and more sustainable society [77]. To manage informational privacy and to build consumer trust in AVs, scholars have recommended increasing transparency regarding what data is being collected, who is using it, how the data is being used and shared, in-depth disclosures about potential security vulnerabilities, providing opt out options, and limiting data collection to the minimal amount required [21,77].

Another realm of privacy that has wide implications for society is surveillance, whereby people can be tracked constantly while they travel [17]. Such a situation is less likely to arise when individuals privately own AVs. However, in the case where AVs are used as a transportation service, public and private agencies will have access to AVs' communication networks. In an extreme scenario, AVs may be used as platforms for widespread surveillance through the use of location tracking, and audio/visual recording of passengers [21]. The idea of widespread surveillance has emerged in China, whose government in 2014 announced plans to launch what they call a 'Social Credit System' by 2020 to rate citizens' financial creditworthiness, and social and political behaviour [84,85]. In this way, the Chinese government aims to restore lost confidence in public institutions by cracking down on corrupt officials and "keep[ing] track of the changing views and interests of China's population" [84].

However, there are concerns that this will undermine individual freedom and democratic processes [12,78,84]. Rannenberg [78] argues that recording personally identifiable data about AV users' geographical locations and destinations, such as participation in an interest group or travelling to a political meeting, could reduce citizens' participation in democratic processes as citizens may assume that such recording could "expose them to risks" and have detrimental impacts on social wellbeing [17]. In a more extreme scenario, Boeglin [86] highlights that using AVs for widespread surveillance has a similar effect as that in a Panopticon, in which individual autonomy is severely restricted due to the "fear of being seen". Thus, AVs' facilitation of widespread surveillance could potentially create a less equitable society, cause greater social unrest, and impede the development of smart governance, whereby citizens have access to and actively utilise information to influence policy-making, which is an important aspect of smart cities as highlighted by De Jong et al. [4] and Caragliu et al. [27].

Thus, having proper governance frameworks to manage the risks to privacy is essential to promote the continued usage of connected infrastructure and information and communication technology (ICT) for socio-economic development, which underscores the advocated model of development for a smart city that requires connectivity by various scholars [4,26,27]. In the next section, we analyse the responses of various governments to managing these privacy risks.

*4.2. Strategies Adopted to Address AV Privacy Risks*

Most governments have begun enacting legislation to manage privacy risks that are not specific to AVs, whereas others, such as Australia, acknowledge the need to address these risks and have begun exploring regulatory options for doing so. The UK and Germany have introduced non-mandatory privacy guidelines that AV manufacturers can follow. An analysis of US states shows that most states have begun introducing and enacting legislation to control AV-specific privacy risks.

The EU has been one of the first entities to focus on privacy and cybersecurity risks. In April 2009 the European Parliament recognised the need to ensure privacy in Intelligent Transport Systems (ITS) from their early stages of the design process, and emphasised the urgency of involving all of the stakeholders in designing these systems [87]. Following a study in October 2012 on how data privacy can be ensured in ITS [88], all EU member states mutually committed to addressing data and cybersecurity issues through the Declaration of Amsterdam on 14 April 2016 [89]. On the same day, the EU followed through on its commitment by successfully ratifying the EU General Data Protection Regulation (EU GDPR), which will take effect in May 2018. The EU GDPR makes significant changes to the Data Protection Directive 95/46/EC of 1995. The regulations even transcend the geographical boundaries of the EU, applying to all companies regardless of their location as long as they process data from people residing in the EU [90]. In the EU GDPR conditions for customers' consent have been strengthened and penalties for violations are more stringent than before as fines are now increased and can be up to 4% of companies' global revenue [90]. Other aspects of the regulations restrict the use of personal data to avoid any potential violation of privacy. However, if these rules are implemented excessively, privacy risks may be avoided at the expense of the benefits of data sharing. This is especially so as AVs rely on a large amount of geo-coded data to navigate optimally [29]. Scholars have highlighted the need to balance privacy concerns with "the utility of the generated information" in accelerating the future improvements to AVs [12], as excessive regulation of the commercial use of such data could undermine the competitive advantage of European vehicle manufacturers [29].

China and Japan have similarly addressed personal data privacy and cybersecurity risks by amending existing legislations. China's new cybersecurity law was enacted in November 2016 and took effect in June 2017. The law requires the anonymisation of all forms of personal information and emphasises on customer consent and transparency from network operators about the purpose, technique, and scope of all data collection and usage. Network operators are also not allowed to "disclose, alter, or destroy collected personal information" unless the person from whom the information was gathered consents, and a person is entitled to request the deletion of their information in the event of illegal or unauthorised collection or use [91]. All the restrictions outlined by the law are not specific to AVs and applies to all data in general. Japan also updated its Privacy Protection Law in December 2016, which took effect in May 2017. The updated law requires organisations to obtain individual consent before using or sharing personal data with third parties, and to notify the public about the reasons for such data usage or data sharing as well as requiring organisations to follow rules set forth by the Personal Information Protection Commission (PIPC) to ensure that it is "impossible to identify a specific individual and restore the personal information" from the anonymous information, and to disclose to the public which "categories of information relating to an individual" are contained in the anonymous information [92]. Scenarios where these requirements do not apply include if the disclosure of the use of data can potentially harm "a principal or third-party's life, body, fortune or other rights and interests" or the data handling organisation's "rights or legitimate interests"; however, it requires organisations to keep records of data transfers to third parties—something the PIPC will provide guidelines on in future [92].

In Singapore, the government is in the process of strengthening regulations on privacy that apply to all personal data. In July 2017, the Personal Data Protection Commission (PDPC) proposed modifications to the Personal Data Protection Act (PDPA) and consulted a range of stakeholders. Key changes that the PDPC has decided to implement include notifying individuals of the purpose behind the collection, use and disclosure of personal data, allowing individuals to opt out from the collection or use of personal data within a "reasonable time period", and requiring organisations to conduct "risk and impact" assessments to ensure that the usage of data is unlikely to negatively impact individuals [93]. Similar to Japan's amended Privacy Protection Law, Singapore's PDPC also intends to make an exception to these rules if the data is being used to protect "legitimate interests" that have "economic, social, security or other benefits for the public", while still requiring full disclosure of the reasons for adopting this exception and is planning to release guidelines to clarify the definition of "adverse impact" and "legitimate interests", as well as situations where opt outs are not required [93]. The Singapore government has also introduced a new law in November 2017 that regulates data sharing specifically between public sector agencies. The Public Sector (Governance) Bill establishes a new data sharing framework, which imposes criminal penalties for unauthorised data sharing between public agencies, and unauthorised re-identification of anonymised data [94,95]. The Bill seeks to improve the "efficiency or effectiveness of policies, programme management or service planning and delivery by Singapore public sector agencies" and to support a "whole-of-government approach" to delivering public services [95] and strengthen the security of sensitive information, such as identifiable data, through storage in databases additional protection measure and limited access [94]. Similar to the EU, China and Japan, the Public Sector (Governance) Bill establishes specific boundaries around the sharing and use of data by public sector agencies, which might alleviate concerns around the use of data for surveillance. Furthermore, it lists the penalties that will be applied in a variety of scenarios where data protection is violated. For instance, it imposes a fine of S$5000 or imprisonment of up to two years if a public sector official attempts to re-identify anonymised data without authorisation [95].

The South Korean and US governments have taken legislative actions to address the data privacy risks applicable to all vehicles. In 2017, the US introduced the SPY Car Act, which regulates access to and use of data stored in all vehicles manufactured for sale in the US [96]. Firstly, the law mandates that owners or lessees must be clearly informed regarding the "collection, transmission, retention, and use" of the data collected from the vehicle. Secondly, the law mandates certain provisions to ensure consumer control over privacy. For instance, users of the vehicles must be able to terminate the collection and retention of data in the vehicle, and in the event of doing so, they should still have access to all features and services provided by the vehicle. Thirdly, the law attempts to limit the misuse of personal data for commercial purposes by explicitly stating that unless vehicle users have given their consent, manufacturers are not allowed to use such data for marketing and advertising purposes. Similarly, in 2016 South Korea updated its Vehicle

Management Act and established provisions for AV testing, as well as data collection involving all vehicles; the Minister of Land, Infrastructure and Transport must grant permission before any individual can use collected data. The Act focuses on the avoidance of privacy violations in respect of vehicle owners and demands the approval and determination by the Ministry, without specifying the level of information sharing that is permitted [97,98].

The governments in Germany, France and Australia have acknowledged the importance of AV-related privacy concerns but have yet to take legislative action to manage these risks. The German government's AV legislation that was enacted in 2017 addresses mainly liability risks but does not address privacy risks [99]. For instance, the law requires a black box (to record the AV journey) for apportioning liability during accidents, but it does not clarify the ownership and sharing of the data collected in the black box. However, the intention is to revise the bill in two years to incorporate AV data use and protections into the law [99]. The government in France has not amended legislation relating to privacy in general but intends to revise regulations for AVs by 2019 to legalise the testing of level-four AVs on public roads in France [100]. In addition, the government intends to allow for greater data sharing to help French companies and researchers further develop AI including AVs while committing to incorporate privacy concerns in a manner consistent with the EU's GDPR [100].

Australia has also released non-mandatory privacy recommendations and, in addition, has consulted different stakeholders to further explore the implications of privacy and ways in which privacy risks can be addressed. The National Transport Commission (NTC) has advocated a "privacy by design" approach to AVs and the incorporation of privacy protection at the highest level, focusing on not generating personal information in the first place [101]. Another consultation was issued by the House of Representatives Standing Committee on Industry, Innovation, Science and Resources (HRSCIISR) to explore the social implications of AVs. The HRSCIISR reported the need to further examine data rights for a range of stakeholder groups and recommended establishing a national body, together with state and territory jurisdictions, and in collaboration with AV manufacturers and technology providers, to coordinate the introduction of AVs into the marketplace [102]. Australia currently relies on existing privacy laws, the Privacy Act 1988, and the Australian Privacy Principles, to regulate manufacturers and technology providers. The NTC perceives these existing provisions to be sufficient in the short run, explicitly stating that "no changes are recommended at this time to privacy laws governing automated vehicles and the transmission of personal information (including location data)" and plans on developing options for sufficiently protecting privacy for AV users, and regulating government access to AV data by late 2018 [101]. The NTC recognises that current restrictions to government access to data specified in the Privacy Act 1988 are not very strong as the Australian Privacy Principles are based on a consent model while consumers are often unaware of how and what data is collected and used [101].

In the UK, the Department for Transport (DfT) and the Centre for the Protection of National Infrastructure (CPNI) collaborated and released eight key principles for privacy and cybersecurity in AVs in August 2017. These principles are also non-mandatory and serve as guidelines that AV manufacturers and other parties in the supply chain should follow. The principles recommend that manufacturers follow the ISO 27018 and ISO 29101, which outline best practices in relation to protecting personally identifiable information (PII) in public clouds, and provide a Privacy Architecture framework, respectively [103]. While most of the principles emphasise the cybersecurity of the AV systems, privacy considerations are included in some aspects. For instance, they recommend "minimising shared data storage" and briefly state that PII must be "managed appropriately" in terms of data storage, transmission, and usage, as well as data ownership [103]. In addition, they recommend that users should have the option to delete "sensitive data" collected in AV systems. These privacy principles are similar to the EU GDPR and the SPY Car Act in the US. However, they do not specify the scenarios in which PII collection and usage is prohibited or permitted. For instance, how PII should be "managed appropriately" and "sensitive data" is not defined. Overall, the guidelines indicate the UK government's awareness of privacy risks relating to AVs, but they have not made these guidelines mandatory. A possible reason for this is the goal of making the UK a leader for AV R&D [104], and thus the government is careful not to inhibit the growth of the technology. The German government has also pursued a similar approach to the UK and 13 recommendations for AVs have been released that mainly address safety risks, but also include recommendations for greater clarification regarding "which data businesses can process without the 'explicit consent' of vehicle users" [105].

At the state level in the US, most surveyed state governments have so far amended or introduced legislation to tackle AV-related privacy risks. A common requirement in state legislation is for AV testers to make full disclosure to passengers regarding what personal information is collected, and how it is collected and used. California was the first to impose this requirement, in 2012 [106]. In late 2015, California's Department of Motor Vehicles (DMV) released draft requirements for manufacturers to disclose to the operator if information beyond what is necessary for safety is collected from the AV, and manufacturers are required to obtain approval before collecting this information [107]. These draft requirements have since become part of new, permanent legislation and manufacturers must fully disclose the collection of any information that is not necessary for ensuring the safety of the driver, and the information must be anonymised [108]. Other states following suit are Georgia [109], Massachusetts [110], Michigan [111], and Tennessee [112]. Since 2016, these states have introduced the Safe Autonomous Vehicles (SAVE) project, which allows eligible motor vehicle manufacturers to deploy a network of on-demand AVs. Any individual participating in the SAVE project is deemed to consent to data collection before and during the project, and manufacturers must also release a privacy statement to the public disclosing its data handling practices related to the AV fleet [109–112]. In March 2017, Texas introduced a bill to manage privacy risks from AVs offering transportation services [113], specifying similar disclosure requirements to those of Georgia, Michigan, Massachusetts, and Tennessee regarding the SAVE project. However, the bill failed to pass. In April 2017, Tennessee updated its data breach notification law, which revised definitions of a "breach of system security" and "personal information" to include the specification that such a breach "materially compromises the security, confidentiality, or integrity of personal information" of the information holder [114].

Massachusetts has taken a broader approach by introducing a new bill on January 2017 to regulate the data collected by Internet of things (IoT) devices, which includes AVs [115]. These regulations explicitly apply to anyone who owns or licenses personal information, collects personal data, or manufactures any AV that uses or installs an IoT device in the vehicle that collects personal data. The aim is to ensure the confidentiality and security of the customer information "in a manner fully consistent with industry standards", to guard against "anticipated threats or hazards to the security or integrity of such information" and "against unauthorised access to or use of such information that may result in substantial harm or inconvenience to any consumer" [115]. Another bill proposed in January 2017 instructs the Massachusetts Department of Transportation and the Registrar of motor vehicles to create rules for the testing, deployment, sale, and leasing of AVs [116]. The public can openly access data collected from AVs, but subject to safeguards as deemed necessary by the Registrar for privacy protection and the bill limits the storage of safety data to a maximum of 18 months which may represent an attempt to limit the potential for data misuse. Pennsylvania has not enacted new legislation regarding data privacy but has made recommendations regarding data collection and usage. Pennsylvania Department of Transportation (PennDOT) released draft rules in 2016 that require all AV testers to provide information collected in the vehicle to the department [117]. With some exceptions, the department declared its commitment to non-disclosure of confidential information and in fact defined the term (PennDOT defines confidential information as "knowledge, information, data, compilations of data, customer-identifying information, reports, and documents that are confidential, trade secrets of, and proprietary to the Tester (i.e., information not in the public domain) including, but not limited to, information about and on the Tester's products, customers, and business operations and strategy" [117]), ensuring that exempted third parties will keep the data confidential. PennDOT has not, however, specified the steps it will take to ensure the confidentiality of such data by third parties.

For safety reasons, manufacturers and AV testers are permitted to collect data on the total number of miles travelled by, or hours of operation of, AVs, the extent of AV fleet testing, and the number of reportable crashes where the AV is deemed to be at fault as well as other types of data such as the number of new jobs created in Pennsylvania due to AV testing [117].

## 5. AV Cybersecurity

### 5.1. AV Cybersecurity Implications for Smart and Sustainable Cities

The cybersecurity of AVs is essential for smart and sustainable development. The term "security" in cybersecurity is often defined as the protection of the integrity, confidentiality, and accessibility of information [77,118]. Von Solms and van Niekerk [119] highlight that the requirements of cybersecurity extend beyond this definition, which fits more closely to that of informational security which seeks to protect the underlying ICT infrastructure, and information that is both directly and indirectly stored and transmitted by the technology [119]. However, in the case of cybersecurity, all assets (both information-based and non-information-based) need to be protected to secure not just the cyberspace, but also "those that function in cyberspace, whether individuals, organisations or nations" as the non-information-based assets such as the personal and physical aspects of human beings, societal values, and national infrastructure also need to be protected [119]. For this reason, cybersecurity concerns encompass both privacy and safety [76,77].

Firstly, the cybersecurity of AVs is essential for safety and social stability, one of the key criterions for a sustainable city. The vulnerabilities facing cybersecurity in general stem from the use of ICTs and their interaction with cyberspace [119]. Given that AVs require external communication networks to operate safely on roads, there is a risk of third parties hacking these wireless networks and undermining safety-critical functions of the AV systems [10,29,120]. Scholars have highlighted various ways in which AVs can be hacked and their negative implications, such as hacking the AVs' wireless Event Data Recorder system [12], jamming the AVs' GPS signal for the purposes of theft [11,121], modifying the AVs' sensors and maps to distort perceptions, and conducting Denial of Service (DoS) attacks to prevent receiving critical information [10,11,121,122]. Yağdereli et al. [122] highlight the security weaknesses of the controller area network protocol that interconnects a broad range of functionalities in modern vehicles, such as its vulnerability to malicious content in any one component in the network. Given AVs' ability to store a variety of private information, such as credit card usage and medical records [120], hackers have great incentives to obtain this information to commit crimes. The cybersecurity of AVs also has wider societal implications beyond road safety as AVs form part of a nation's critical infrastructure. Iacono and Marrone [123] highlight that attacks against critical infrastructure are increasing as cyber-physical systems, such as AVs, expose critical infrastructure systems to the Internet and its related cyber threats. This can compromise the integrity and availability of data and access to the critical services provided by the infrastructure, such as resulting in accidents and traffic disruptions, thereby affecting the society's wellbeing as a whole [83,119].

Secondly, the cybersecurity of AVs is essential for economic sustainability as it enables the realisation of the gains from improved connectivity. Cyber-attacks, in general, are costly for businesses, and it is estimated that cyber-attacks will cost approximately $3 trillion in lost productivity by 2020; therefore, enhancing cybersecurity can help save on costs and build firms' economic competitiveness in the longer-run [13]. Given the aforementioned implications of AVs' cybersecurity for safety, cybersecurity is crucial to building trust in, and increasing public acceptance of, AVs [77]. In AVs, cyber-attacks violate the expected confidence in computer systems and thus erode user confidence in the technology. Numerous studies have provided recommendations on what kinds of robust security and privacy frameworks can be adopted to improve cybersecurity and to increase users' trust in technology in general [13,124]. Promoting cybersecurity through the lens of corporate social responsibility can also allow firms to "distinguish themselves and add value", which can improve business outcomes in the long run [13]. Unless AV developers establish sound cybersecurity practices for the technology, cyber-attacks resulting from

systematic weaknesses will cause a "critical setback to the connectivity efforts and progress made over the last years" [124]. Currently, the automotive industry standards lack sufficient coverage for the breadth of cybersecurity risks faced by AVs and scholars highlight the need for a "solid foundation of security, privacy, and trust to fully take advantage of innovations in connected and self-driving vehicles", which would require the collaboration between a diverse group of stakeholders to ensure both the physical and cyber safety of AV users [77].

*5.2. Strategies Adopted to Address AV Cybersecurity Risks*

The commonly adopted approaches to address cybersecurity risks include enacting legislation that are not specific to AVs, conducting further research into cybersecurity risks for AVs and all vehicles in general, and providing guidelines. Some governments such as that of the UK and Singapore have also begun educating the public of cybersecurity risks, whereas the governments of Japan and South Korea have yet to indicate their intentions to address cybersecurity risks.

In the US, the federal government has taken steps to explore vehicle cybersecurity risks and has made recommendations to manage AV-specific cybersecurity risks. In 2012, the National Highway Traffic Safety Administration (NHTSA) set up a new department to research the "safety, security, and reliability of complex, interconnected, electronic vehicle systems" and set up an Electronics Council to enhance collaboration across the entire NHTSA organisation regarding vehicle electronics and cybersecurity [125]. More recently, the NHTSA, released non-mandatory recommendations for AV development and in the Automated Driving Systems guidance, encouraged entities to design their AV systems according to standards established by relevant organisations, such as NHTSA, SAE, the Automotive Information Sharing and Analysis Centre (ISAC), and various auto-manufacturer associations [126]. Other recommendations include development of cyber incident response plans and vulnerability disclosure and reporting policies as well as publishing a Voluntary Safety Assessment letter to demonstrate that entities are following the recommended guidelines; However, surprisingly given the ample emphasis on developing plans, policies and guidelines, the NHTSA also explicitly states that to avoid delays in testing or deployment, submission of the Assessment Letter is "not required", and there is no "mechanism to compel entities to do so" [126], which may stem from the fear of stifling AV developments.

In the US, the recently introduced SPY Car Act addresses vehicle cybersecurity risks. The law contains provisions to guard against the hacking of vehicles, such as requiring penetration testing to evaluate vehicles' resilience to hacking and separating critical and non-critical software systems in all vehicles, and it provides specifications for ensuring the security of the data collected in vehicles and during transmission from the vehicle, and of data stored outside the vehicle [96]. The SPY Car Act requires AVs to have the capability to detect, prevent, and report attempts at hijacking the control of vehicles, as well as capturing the stored data.

At the state level, California, Massachusetts, and Pennsylvania have introduced legislation to address cybersecurity threats to AVs. California's DMV drafted rules in 2016 that require manufacturers to certify AVs' ability to detect and react to cyber-attacks in accordance with the "appropriate and applicable" industry standards, which has since been approved as permanent regulation on February 2018 [108]. Massachusetts' newly-introduced bill authorises the Department of Consumer Affairs and Business Regulation to implement regulations that are consistent with federal regulations to protect personal information and data collected by an IoT device, which includes AVs and other smart devices [115]. The delegation of additional responsibility to the Department is done to guard against unauthorised access or use and protect the integrity of the data. Another introduced bill assigns responsibility to the Massachusetts Department of Transportation to ensure that data collection systems are secure, tamper-resistant, and able to maintain the accuracy of the data collected [127]. Unlike California and Massachusetts, Pennsylvania's new bill does not include specifications to ensure the security of data collection systems but makes non-mandatory recommendations for AV testers to provide proof that cybersecurity precautions are taken, and it imposes requirements for AV testers and PennDOT to notify each other immediately in the event that a cybersecurity intrusion attempts to access connected infrastructure or the AV [117,128].

Georgia, Michigan, and Texas have enacted legislation that addresses cybersecurity risks to all systems in general, although these are not specific to AVs. Georgia recently passed a new bill that criminalises unauthorised access to computers or computer networks in the state [129]. In 2017, Texas authorised a new Cybersecurity Act that instructs the creation of a committee to research on cybersecurity issues and the "information security plans" of Texas' state agencies [130]. In the same year, Michigan enacted a law that allocates funds for cybersecurity and instructs the Michigan DOT to "identify specific outcomes and performance measures" to improve cybersecurity, such as by reducing the daily incidence of cyber threats, and to raise awareness among citizens about cyber threats and the necessary preventive actions [131]. The government of Michigan also proposed creating a cybersecurity council to recommend cybersecurity improvements to state infrastructure and to identify ways to enhance the state's cybersecurity industry [132].

The EU has adopted a variety of strategies to manage cybersecurity risks. In August 2016, the EU enacted the first EU-wide legislation on cybersecurity: "the Directive on the security of network and information systems" (NIS directive) [133]. In August 2014, the EU's Data Protection Working Party highlighted the potential security risks of IoT [87] and in December 2016 the EU Agency for Network and Information Security released best practices guidelines for the cybersecurity of connected vehicles, including both conventional and AVs, to increase awareness of and provide guidance on these issues [134].

China has also taken steps to address the cybersecurity risks of all cyber systems. China's new cybersecurity law takes effect in June 2017, outlining specific provisions to strengthen the protection of critical infrastructure and personal information, as well as establishing the responsibilities of network operators, regulating the sale of critical cybersecurity equipment, and outlining penalties for potential violations [135]. Network operators are required to follow security procedures to "safeguard networks from interference, destruction or unauthorised access" [136]. These stricter requirements will shape how foreign businesses handle data in China, as the Cyberspace Administration of China (CAC) also published "Measures for Security Assessment of Personal Information and Important Data Leaving the Country" earlier in April 2017, expanding the requirements for data localisation to all network operators which will impact AV manufacturers looking to test and deploy their products in China [91]. Many provisions of the law appear to be designed to protect national interests. For instance, the law emphasises on ensuring the security of data moving in and out of China [136], and consistently evaluates the security of critical information infrastructure that is vital for the public interest as well as the national economy and national security [135]. The law also requires sensitive information to be preserved in the country, although the government has yet to explicitly clarify what kinds of information they deem to be "sensitive" [136].

Singapore has adopted a diversified approach to managing cybersecurity risks by amending legislation, consulting different stakeholders, and educating the public about such risks. In April 2017, the Singapore government amended the Computer Misuse and Cybersecurity (CMC) Act, which took effect in June 2017, making it illegal for individuals to use personal information "obtained illegally from a computer" and to obtain "hacking tools" to commit or facilitate crimes [137]. The government has taken steps to strengthen the response to such risks and minimise the costs associated with these risks. The PDPC issued a public consultation in July 2017 and proposed a "mandatory data breach notification regime" whereby organisations are required to notify affected individuals and the PDPC in the event of a data breach that is to cause significant impact on individuals to whom the breached information relates [93]. In addition, exceptions are made if the affected individuals are the subjects of an "ongoing or potential investigation under the law" and if the data being unauthorised for use or collection is encrypted [93]. Overall, the PDPC's proposed changes aim to balance the need for using data with individuals' rights to privacy protection.

On the other hand, the governments of Japan and South Korea have not provided any indication of their approach to managing cybersecurity risks in relation to AVs or cyber systems as a whole. In 2017, the Korean government amended the Motor Vehicle Management Act. However, the law does not include any provisions on cybersecurity. The law states that any person who "intends to use the data processed by electronic data processing systems" required for management of vehicles must obtain approval from the Minister of Land, Transport and Maritime Affairs [97], but it does not provide any information on, for instance, whether there are any restrictions on data access or any measures to prevent unauthorised access to such data. Similarly, Japan has not amended legislation or provided guidelines on addressing cybersecurity risks in general or those that are specific to AVs.

While governments of Germany, France, Australia, and the UK have not amended or introduced any new legislation on cybersecurity, they have taken steps to increase awareness of AV-related cybersecurity risks. The German government has established working groups on AV-related issues in September 2015, which include cybersecurity and data protection as part of its national strategy for "Automated and Connected Driving" [138]. Likewise, the French government has set up working groups in 2016 to address AV-related societal issues, one of which involves security issues. As mentioned in the Privacy section, in Australia, the HRSCIISR has recommended coordination of the efforts of different stakeholders through creation of a national body [102].

The UK's approach to addressing cybersecurity risks seems to be targeted at increasing awareness, strengthening the nation's longer-term resilience to cybersecurity risks and building the nation's cybersecurity industry that includes focus on AVs [139]. The UK tech consortium, 5*StarS, secured funding from the government's innovation agency, "Innovate UK", in April 2017 to develop a methodology that the AV industry can use to ensure that AV systems meet the required cybersecurity standards throughout their life span [140]. The UK has also implemented the National Cybersecurity Strategy 2016–2021 that focuses on promoting research and strengthening UK's position in this field by 2021. The government has also set up the National Cybersecurity Centre (NCSC) in October 2016 to provide guidance on addressing cybersecurity risks, as well as to aid in the implementation of the EU's NIS directive [141]. As mentioned in the previous section, the UK DfT and CPNI have created a set of principles that are aimed at the AV manufacturing supply chain. Principles 2 and 5–8 relate to key aspects of AV manufacturing ranging from focusing on security in the design process and resilience and response to attacks (Principles 5 and 8), lifecycle management of AV software systems and management of risks in the supply chain (Principles 6 and 2) to security in the storage and transmission of data (Principle 7) [103]. Figure 1 summarises the establishment of AV-related privacy and cybersecurity strategies by different governments discussed in Sections 4 and 5.

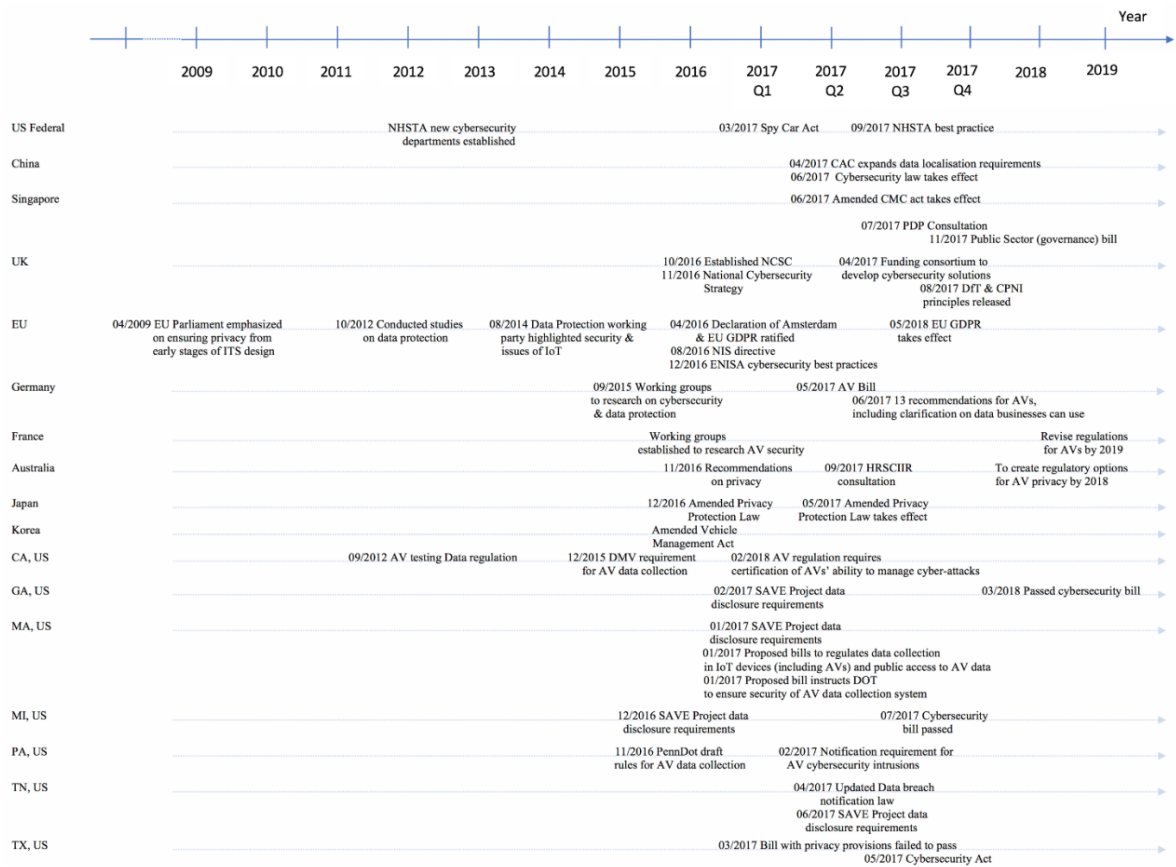

**Figure 1.** Timeline of AV-related privacy and cybersecurity strategies.

## 6. Discussion and Conclusions

Transportation has immense implications for social welfare, economic development, and environmental sustainability. Congestion, environmental degradation, social inequity, and public health issues are problems that sustainable transport policies urgently need to resolve. In this article, we explored AVs and their potential as a solution for smart and sustainable development. AVs offer potentially transformative benefits that can alleviate some of these concerns and lead the way to a greater level of sustainability. Due to AVs' connected nature, data can be used to optimise traffic flows, reduce congestion, and yield significant economic benefits, as well as increase accessibility by new user groups. The increased accessibility can lower psychological barriers to mobility, and isolation, and thus improve public health and wellbeing. These economic and social benefits can be realised while ensuring the self-sufficiency of the environmental system, as it has been shown that AVs can reduce carbon emissions and fuel consumption. Studies point out that these benefits may present a trade-off with other unintended consequences. Greater accessibility to mobility may create induced demand that can worsen congestion, increase emissions, increase energy consumption, and cause a deterioration in public health if individuals increase AV usage at the expense of more active modes of transport, which will worsen the effects of a sedentary lifestyle. Furthermore, the benefits to congestion depend on many factors, such as the penetration rate of AVs and the increase in road capacity.

As highlighted in this article, AV privacy has crucial implications for smart and sustainable development. Amidst the growth of ICTs and the sharing economy, the protection of personal data and the security of communication networks are vital to ensure society capitalises on the gains from increased connectivity. Firstly, our research shows that informational privacy is crucial to protect AV users against the misuse of data that can exacerbate existing social inequalities. Secondly, informational privacy is important to build consumer trust, which is required if further developments and improvements are to be made for AVs' longer-run success. Thirdly, AV privacy is important to guard against widespread surveillance, as the latter can potentially undermine democratic processes and discourage citizens' active usage of connected platforms, which in turn can be detrimental to social equity and impedes the development of a smart city.

Most of the surveyed governments have responded to privacy risks in general by enacting new legislation that applies to all personal data. The US SPY Car Act represents an attempt to monitor or manage the agencies in control of the information collected in vehicles, such as restricting data usage for advertising or marketing purposes. However, the Act has yet to introduce similar restrictions on data usage by insurance companies to determine insurance premiums, which have significant economic and social implications. The EU has explored privacy and cybersecurity risks to all cyber systems in general much earlier and its GDPR appears more restrictive relative to other countries in breadth and depth, prohibiting any automated decision that compromises its citizens' privacy. It has been suggested that greater data sharing in Europe is impeded by the perception that European markets are "purely domestic" and that data sharing undermines Europe's competitive advantage [142]. While proper implementation of the GDPR can ensure privacy protection, it is essential also to monitor the developments in other countries that face fewer data restrictions. The UK's recommendations for privacy are similar to the provisions of the EU GDPR but are less specific in defining the ways in which data collection and usage is permitted. The French government acknowledges privacy concerns but is aware of the importance of greater data sharing for the development of AVs and AI in general, which the French government aims to further enhance to "rival" that in the US and China [142]. In particular, data privacy is much less of a concern in China compared to the EU and US, which be advantageous as there are fewer impediments to the storage, collection and use of data [142,143]. The Singapore government is amidst updating its privacy laws to further clarify exemptions to such restrictions to balance privacy needs with the public benefits that arise from using personal data. On the other hand, states in the US have generally enacted or introduced regulation to manage privacy risks specific to AVs. Voluntary guidelines on privacy best practices were released by Germany and Australia but in the case of Germany act as a temporary measure before privacy concerns are incorporated into legislation in future.

The cybersecurity of AVs is another key factor for sustainable development in cities. With the proliferation of ICTs and society's increasing dependence on connected platforms, such as AVs, the security of both the information being stored on such platforms and the individuals to whom the information relates is crucial for both social stability and economic sustainability. Our research shows that AVs are especially vulnerable to cyber-attacks due to their ability to store highly sensitive data and transmit such data on external communication networks. These networks can be hacked in a multitude of ways for committing crimes or undermining the safety of the AV, which has tremendous impacts on road safety and social stability. Insufficient cybersecurity of AVs can also expose a nation's critical infrastructure to cyber threats that can disrupt the delivery of critical services and have a detrimental impact on the entire society's wellbeing. Sufficient market uptake of AVs is required for further improvements to be made to the technology to realise the full scope of its economic, social, and environmental benefits. As such, cybersecurity of AVs is crucial to build trust in and increase public acceptance of AVs, which in turn can help increase its rate of adoption.

The responses of governments to address cybersecurity risks appear more varied. Some governments have enacted or amended legislation that applies to all cyber systems, such as in China and Singapore. China's new law emphasises on strengthening the security of critical infrastructure and networks, whereas Singapore's amendments to its cybersecurity laws appear more targeted at strengthening the public's response to data breaches. It has been noted that the presence of multiple regulatory agencies in China to oversee the implementation of these rules may complicate the initial stages of enforcement [144]. In comparison to the steps taken by other countries to address cybersecurity risks, China's new cybersecurity law places a large emphasis on the protection of national sovereignty by, for instance, requiring all sensitive data to remain within China. This may negatively impact foreign AV manufacturers looking to test AVs in China, as it is likely that foreign AV developers are not permitted to "engage in mapmaking and comprehensive data collection on Chinese soil" [143]. Nevertheless, the Chinese government is "clearly committed to helping Chinese companies lead the world" AV technology as it has identified AVs as one of its key priorities in its "Made in China 2025" initiative to lead the world in innovation. Its commitment is demonstrated in its official policies, which encourage Chinese technology companies to cooperate in devising transportation solutions including AVs [143]. State governments in the US have enacted legislation to address cybersecurity risks in specifically AVs, such as California's regulations and Massachusetts' new law that directs authorities to regulate personal data collected in all IoT devices. At the national level, the US government's SPY Car Act address cybersecurity risks in all vehicles via requirements to ensure the robustness of vehicles and their networks against cyber-attacks. Some governments have pursued diversified approaches to manage cybersecurity risks. For instance, Singapore has amended existing legislation but has also issued public consultations for future regulatory changes, the EU has enacted the NIS directive while also releasing voluntary cybersecurity guidelines, and the UK has rolled out a national cybersecurity strategy, which includes investing in nation's cybersecurity industry and strengthening its citizens' cybersecurity skills. Some governments have mainly focused on conducting research on AV cybersecurity risks, such as that of Germany, France, and Australia, while Japan and South Korea have yet to indicate their approach to address cybersecurity threats.

This study serves to inform policy-makers, scholars, and various stakeholders in the automotive industry of privacy and cybersecurity challenges of AVs for achieving smart and sustainable cities and highlighted in-depth the emerging strategies to address these concerns by different governments. It should be noted that in our study we examined the relation between AVs and smart and sustainable development and highlighted the existing steps taken to address privacy and cybersecurity risks in AVs. A myriad of factors affect how each government addresses different types of risks, and these strategies change over time. As such, a comprehensive analysis of each country is required in future to sufficiently justify the reasons behind governments' strategies to privacy and cybersecurity concerns and changes to these strategies over time. The authors plan to use single and comparative case studies in future to address this question appropriately and hope that this study encourages other researchers to study the implications of AVs for smart and sustainable development.

## References

1. United Nations. *Sustainable Development Goals*; United Nations: New York, NY, USA, 2015.
2. INRIX. *Americans Will Waste $2.8 Trillion on Traffic by 2030 if Gridlock Persists*; INRIX: Kirkland, WA, USA, 2014.
3. Pangbourne, K.; Stead, D.; Mladenović, M.; Milakis, D. The case of mobility as a service: A critical reflection on challenges for urban transport and mobility governance. In *Governance of the Smart Mobility Transition*; Emerald Publishing Limited: Bingley, UK, 2018; pp. 33–48.
4. De Jong, M.; Joss, S.; Schraven, D.; Zhan, C.; Weijnen, M. Sustainable–smart–resilient–low carbon–eco–knowledge cities; making sense of a multitude of concepts promoting sustainable urbanisation. *J. Clean. Prod.* **2015**, *109*, 25–38.
5. Gopalakrishnan, K.; Chitturi, M.V.; Prentkovskis, O. Smart and sustainable transport: Short review of the special issue. Taylor & Francis: Abingdon, UK, 2015.
6. Park, J. Sensemaking/What Will Autonomous Vehicles Mean for Sustainability? Available online: https://thefuturescentre.org/articles/11010/what-will-autonomous-vehicles-mean-for-sustainability (accessed on 10 January 2018).
7. Allianz. Autonomous cars in smart cities. *The Telegraph*, 9 October 2017.
8. Lubell, S. Here's how self-driving cars will transform your city. *Wired*, 21 October 2016.
9. Parkin, J.; Clark, B.; Clayton, W.; Ricci, M.; Parkhurst, G. Autonomous vehicle interactions in the urban street environment: A research agenda. *Proc. Inst. Civ. Eng. Munic. Eng.* **2018**, *171*, 15–25.
10. Petit, J.; Shladover, S.E. Potential cyberattacks on automated vehicles. *IEEE Trans. Intell. Transp. Syst.* **2015**, *16*, 546–556.

11.  Dominic, D.; Chhawri, S.; Eustice, R.M.; Ma, D.; Weimerskirch, A. Risk assessment for cooperative automated driving. In Proceedings of the 2nd ACM Workshop on Cyber-Physical Systems Security and Privacy, Vienna, Austria, 28 October 2016; ACM: New York, NY, USA, 2016; pp. 47–58.

12.  Kohler, W.J.; Colbert-Taylor, A. Current law and potential legal issues pertaining to automated, autonomous and connected vehicles. *St. Clara Comput. High Technol. Law J.* **2014**, *31*, 99.

13.  Shackelford, S.J.; Fort, T.L.; Charoen, D. Sustainable cybersecurity: Applying lessons from the green movement to managing Cyber Attacks. *Univ. Ill. Law Rev.* **2016**, *2016*, doi:10.2139/ssrn.2324620.

14.  Kalra, N. *Challenges and Approaches to Realizing Autonomous Vehicle Safety*; RAND Corporation: Santa Monica, CA, USA, 2017.

15.  Kalra, N.; Paddock, S.M. Driving to safety: How many miles of driving would it take to demonstrate autonomous vehicle reliability? *Transp. Res. Part A Policy Pract.* **2016**, *94*, 182–193.

16.  Brodsky, J.S. Autonomous vehicle regulation: How an uncertain legal landscape may hit the brakes on self-driving cars. *Berkeley Technol. Law J.* **2016**, *31*, 851, doi:10.15779/Z38JC5S.

17.  Glancy, D.J. Privacy in autonomous vehicles. *St. Clara Law Rev.* **2012**, *52*, 1171–1239.

18.  Fagnant, D.J.; Kockelman, K.M.; Bansal, P. Operations of shared autonomous vehicle fleet for Austin, texas, market. *Transp. Res. Rec. J. Transp. Res. Board* **2015**, *2536*, 98–106.

19.  Maurer, M.; Gerdes, J.C.; Lenz, B.; Winner, H.; *Autonomous Driving: Technical, Legal and Social Aspects*; Springer: Berlin/Heidelberg, Germany, 2016.

20.  Meyer, G.; Beiker, S. *Road Vehicle Automation*; Springer: Berlin/Heidelberg, Germany, 2014.

21.  Schoonmaker, J. Proactive privacy for a driverless age. *Inf. Commun. Technol. Law* **2016**, *25*, 96–128.

22.  Brundtland, G.; Khalid, M.; Agnelli, S.; Al-Athel, S.; Chidzero, B.; Fadika, L.; Hauff, V.; Lang, I.; Shijun, M.; de Botero, M.M. *Our Common Future*; Brundtland Report; Oxford University Press: Oxford, UK, 1987.

23.  Noy, K.; Givoni, M. Is 'smart mobility'sustainable? Examining the views and beliefs of transport's technological entrepreneurs. *Sustainability* **2018**, *10*, 422, doi:10.3390/su10020422.

24.  Meadows, D.H. Indicators and information systems for sustainable development. In *The Earthscan Reader in Sustainable Cities*; Earthscan: London, UK, 1998.

25.  Rode, P.; Burdett, R. Cities: Investing in energy and resource efficiency. In *Towards a Green Economy: Pathways to Sustainable Development and Poverty Eradication*; United Nations Environment Programme: Nairobi, Kenya, 2011.

26.  Zawieska, J.; Pieriegud, J. Smart city as a tool for sustainable mobility and transport decarbonisation. *Transp. Policy* **2018**, *63*, 39–50.

27.  Caragliu, A.; Del Bo, C.; Nijkamp, P. Smart cities in Europe. *J. Urban Technol.* **2011**, *18*, 65–82.

28.  Collingwood, L. Privacy implications and liability issues of autonomous vehicles. *Inf. Commun. Technol. Law* **2017**, *26*, 32–45.

29.  West, D.M. *Moving Forward: Self-Driving Vehicles in China, Europe, Japan, Korea, and the United States*; Report; Brookings Institution: Washington, DC, USA, 2016.

30.  Hörl, S.; Ciari, F.; Axhausen, K.W. Recent perspectives on the impact of autonomous vehicles. *Arbeitsberichte Verkehrs-und Raumplanung* **2016**, *1216*, doi:10.13140/RG.2.2.26690.17609.

31.  General Statistics. Available online: http://www.iihs.org/iihs/topics/t/general-statistics/fatalityfacts/state-by-state-overview/2015 (accessed on 2 December 2017).

32.  Thomopoulos, N.; Givoni, M. The autonomous car—A blessing or a curse for the future of low carbon mobility? An exploration of likely vs. Desirable outcomes. *Eur. J. Futures Res.* **2015**, *3*, 14, doi:10.1007/s40309-015-0071-z.

33.  Doecke, S.; Grant, A.; Anderson, R.W. The real-world safety potential of connected vehicle technology. *Traffic Inj. Prev.* **2015**, *16*, S31–S35.

34.  Mouhagir, H.; Talj, R.; Cherfaoui, V.; Aioun, F.; Guillemard, F. Integrating safety distances with trajectory planning by modifying the occupancy grid for autonomous vehicle navigation. In Proceedings of the 2016 IEEE 19th International Conference on Intelligent Transportation Systems (ITSC), Rio de Janeiro, Brazil, 1–4 November 2016; IEEE: Piscataway, NJ, USA, 2016; pp. 1114–1119.

35.  Rizaldi, A.; Althoff, M. Formalising traffic rules for accountability of autonomous vehicles. In Proceedings of the 2015 IEEE 18th International Conference on Intelligent Transportation Systems (ITSC), Gran Canaria, Spain, 15–18 September, 2015; IEEE: Piscataway, NJ, USA, 2015; pp. 1658–1665.

36.  Neate, R. Ford to build 'high volume' of driverless cars for ride-sharing services. *The Guardian*, 16 August 2016.

37. Korosec, K. Baidu and nvidia to build artificial intelligence platform for self-driving cars. *Forbes*, 1 September 2016.

38. Greimel, H. Japan Inc. steps up autonomous-drive push. *Automotive News*, 6 November 2016.

39. Tan, C.K.; Tham, K.S. Autonomous vehicles, next stop: Singapore. *Journeys* **2014**, *November*, 5–11.

40. Dunne, M.J. China deploys aggressive mandates to take lead in electric vehicles. *Forbes*, 28 February 2017.

41. Milakis, D.; Van Arem, B.; Van Wee, B. Policy and society related implications of automated driving: A review of literature and directions for future research. *J. Intell. Transp. Syst.* **2017**, *21*, 324–348.

42. SAE International. *International Standard J3016: Taxonomy and Definitions for Terms Related to on-Road Motor Vehicle Automated Driving Systems*; SAE International: Warrendale, PA, USA, 2014.

43. Cheney, P. How self-driving cars will ease traffic congestion. *The Globe and Mail*, 2017. Available online: https://www.theglobeandmail.com/globe-drive/culture/commuting/how-self-driving-cars-will-ease-traffic-congestion/article15876882/ (accessed on 2 December 2017).

44. Cooper, J. Driverless cars, hypersonic tunnels, no more traffic lights… Is this the future of travel? *The Telegraph*, 2 June 2017.

45. Newcomb, D. Can self-driving cars kill traffic lights? *PC Magazine*, 1 April 2016.

46. Hardy, B.; Fenner, R.A. Towards the sustainability of road transport through the introduction of AV technology. *Proc. Inst. Civ. Eng.-Eng. Sustain.* **2015**, *168*, 192–203.

47. Yang, K.; Guler, S.I.; Menendez, M. Isolated intersection control for various levels of vehicle technology: Conventional, connected, and automated vehicles. *Transp. Res. Part C Emerg. Technol.* **2016**, *72*, 109–129.

48. Kane, M.; Whitehead, J. How to ride transport disruption—A sustainable framework for future urban mobility. *Aust. Plan.* **2018**, 1–9, doi:10.1080/07293682.2018.1424002.

49. Monteil, J.; Nantes, A.; Billot, R.; Sau, J. Microscopic cooperative traffic flow: Calibration and simulation based on a next generation simulation dataset. *IET Intell. Transp. Syst.* **2014**, *8*, 519–525.

50. INRIX. *Traffic Congestion Cost UK Motorists More Than £30 Billion in 2016*; INRIX: London, UK, 2017.

51. Shladover, S.; Su, D.; Lu, X.-Y. Impacts of cooperative adaptive cruise control on freeway traffic flow. *Transp. Res. Rec. J. Transp. Res. Board* **2012**, *2324*, 63–70.

52. Maciejewski, M.; Bischoff, J. Congestion effects of autonomous taxi fleets. *Transport* **2017**, 1–10, doi:10.3846/16484142.2017.1347827.

53. Metz, D. Future transport technologies for an ageing society: Practice and policy. In *Transport, Travel and Later Life*; Emerald Publishing Limited: Bingley, UK, 2017; pp. 207–220.

54. Taeihagh, A. Crowdsourcing, sharing economies and development. *J. Dev. Soc.* **2017**, *33*, 191–222.

55. Heard, B.R.; Taiebat, M.; Xu, M.; Miller, S.A. Sustainability implications of connected and autonomous vehicles for the food supply chain. *Resour. Conserv. Recycl.* **2018**, *128*, 22–24.

56. Crayton, T.J.; Meier, B.M. Autonomous vehicles: Developing a public health research agenda to frame the future of transportation policy. *J. Transp. Health* **2017**, *6*, 245–252.

57. Pettigrew, S. Why public health should embrace the autonomous car. *Aust. N. Z. J. Public Health* **2017**, *41*, 5–7.

58. Nikitas, A.; Kougias, I.; Alyavina, E.; Njoya Tchouamou, E. How Can Autonomous and Connected Vehicles, Electromobility, BRT, Hyperloop, Shared Use Mobility and Mobility-As-A-Service Shape Transport Futures for the Context of Smart Cities? *Urban Sci.* **2017**, *1*, 36, doi:10.3390/urbansci1040036.

59. Newman, P.; Beatley, T.; Boyer, H. Create sustainable mobility systems. In *Resilient Cities*; Springer: Berlin/Heidelberg, Germany, 2017; pp. 53–87.

60. van Schalkwyk, M.; Mindell, J. Current issues in the impacts of transport on health. *Br. Med. Bull.* **2018**, *125*, 67–77.

61. Bonnefon, J.-F.; Shariff, A.; Rahwan, I. The social dilemma of autonomous vehicles. *Science* **2016**, *352*, 1573–1576.

62. Schlossberg, T. Stuck in traffic, polluting the inside of our cars. *New York Times*, 19 August 2016.

63. Eugensson, A.; Brännström, M.; Frasher, D.; Rothoff, M.; Solyom, S.; Robertsson, A. Environmental, safety legal and societal implications of autonomous driving systems. In Proceedings of the International Technical Conference on the Enhanced Safety of Vehicles (ESV), Seoul, Korea, 27–30 May 2013.

64. Li, Z.; Chitturi, M.V.; Yu, L.; Bill, A.R.; Noyce, D.A. Sustainability effects of next-generation intersection control for autonomous vehicles. *Transport* **2015**, *30*, 342–352.

65. Forrest, A.; Konca, M. *Autonomous Cars and Society*; Worcester Polytechnic Institute: Worcester, MA, USA, 2007.

66. Zhao, J.; Zhao, R.; Wang, G.; Zhang, X. Analysis of fuel economy of autonomous vehicle platoon. In Proceedings of the Fourth International Conference on Transportation Engineering (ICTE) 2013: Safety, Speediness, Intelligence, Low-Carbon, Innovation, Chengdu, China, 19–20 October 2013; pp. 980–986.

67. Wadud, Z.; MacKenzie, D.; Leiby, P. Help or hindrance? The travel, energy and carbon impacts of highly automated vehicles. *Transp. Res. Part A Policy Pract.* **2016**, *86*, 1–18.

68. MacKenzie, D.; Wadud, Z.; Leiby, P. A first order estimate of energy impacts of automated vehicles in the United States. In Proceedings of the Transportation Research Board Annual Meeting, Washington, DC, USA, 12–16 January 2014.

69. Hula, A.; Snapp, L.; Alson, J.; Simon, K. The environmental potential of autonomous vehicles. In *Road Vehicle Automation 4*; Springer: Berlin/Heidelberg, Germany, 2018; pp. 89–95.

70. Pyper, J. Self-driving cars could cut greenhouse gas pollution. *Scientific American*, 15 September 2014.

71. McMahon, J. Big fuel savings from autonomous vehicles. *Forbes*, 17 April 2017.

72. Mersky, A.C.; Samaras, C. Fuel economy testing of autonomous vehicles. *Transp. Res. Part C Emerg. Technol.* **2016**, *65*, 31–48.

73. Khondaker, B.; Kattan, L. Variable speed limit: A microscopic analysis in a connected vehicle environment. *Transp. Res. Part C Emerg. Technol.* **2015**, *58*, 146–159.

74. Bughin, J.; Hazan, E.; Ramaswamy, S.; Chui, M.; Allas, T.; Dahlström, P.; Henke, N.; Trench, M. *Artificial Intelligence—The Next Digital Frontier*; McKinsey Global Institute: Brussels, Belgium, 2017.

75. Dhar, V. Equity, safety, and privacy in the autonomous vehicle era. *Computer* **2016**, *49*, 80–83.

76. López-Lambas, M.E. The socioeconomic impact of the intelligent vehicles: Implementation strategies. In *Intelligent Vehicles*; Elsevier: New York, NY, USA, 2018; pp. 437–453.

77. Pype, P.; Daalderop, G.; Schulz-Kamm, E.; Walters, E.; von Grafenstein, M. Privacy and security in autonomous vehicles. In *Automated Driving*; Springer: Berlin/Heidelberg, Germany, 2017; pp. 17–27.

78. Rannenberg, K. Opportunities and risks associated with collecting and making usable additional data. In *Autonomous Driving*; Springer: Berlin/Heidelberg, Germany, 2016; pp. 497–517.

79. The Economist. What machines can tell from your face. *The Economist*, 9 September 2017.

80. Horvitz, E. *Ai, People, and Society*; American Association for the Advancement of Science: Washington, DC, USA, 2017.

81. De Montjoye, Y.-A.; Hidalgo, C.A.; Verleysen, M.; Blondel, V.D. Unique in the crowd: The privacy bounds of human mobility. *Sci. Rep.* **2013**, *3*, 1376, doi:10.1038/srep01376.

82. Ma, C.Y.; Yau, D.K.; Yip, N.K.; Rao, N.S. Privacy vulnerability of published anonymous mobility traces. *IEEE/ACM Trans. Netw.* **2013**, *21*, 720–733.

83. Fagnant, D.J.; Kockelman, K. Preparing a nation for autonomous vehicles: Opportunities, barriers and policy recommendations. *Transp. Res. Part A Policy Pract.* **2015**, *77*, 167–181.

84. The Economist. China invents the digital totalitarian state. *The Economist*, 17 December 2016.

85. Botsman, R. Big data meets big brother as china moves to rate its citizens. *Wired*, 21 October 2017.

86. Boeglin, J. The costs of self-driving cars: Reconciling freedom and privacy with tort liability in autonomous vehicle regulation. *Yale J. Law Technol.* **2015**, *17*, 171.

87. Pillath, S. *Automated Vehicles in the EU*; PE 573.902; European Parliament: Brussels, Belgium, 2016.

88. Eisses, S. *ITS Action Plan—ITS & Personal Data Protection: Final Report*; European Commission: Brussels, Belgium, 2012.

89. Osborne Clarke. *What Developments Are Driving the Deployment of Connected and Autonomous Vehicles?* Osborne Clarke: London, UK, 2017.

90. European Parliament, Council of the European Union. Regulation (EU) 2016/679 of the European Parliament and of the Council of 27 April 2016 on the Protection of Natural Persons with Regard to the Processing of Personal Data and on the Free Movement of Such Data, and Repealing Directive 95/46/EC (General Data Protection Regulation) (Text with EEA Relevance). Available online: http://eur-lex.europa.eu/legal-content/EN/TXT/PDF/?uri=CELEX:32016R0679&from=EN (accessed on 15 December 2017).

91. Xia, S. *China Cybersecurity and Data Protection Laws: Change Is Coming*; Harris Bricken, LLP: Seattle, WA, USA, 2017.

92. PIPC. *Amended Act on the Protection of Personal Information (Tentative Translation)*; Personal Information Protection Commission (PIPC): Tokyo, Japan, 2016.

93. Personal Data Protection Commission. *Response to Feedback on the Public Consultation on Approaches to Managing Personal Data in the Digital Economy*; Personal Data Protection Commission: Singapore, 2018.

94. Seow, J. Parliament: New laws on data sharing between public sector agencies. *The Straits Times*, 8 January 2018.

95. Singapore Status Online. *Public Sector (Governance) Act 2018*; Singapore Status Online: Singapore, 2017.

96. S.680—SPY Car Act of 2017. 115th Congress. United States of America, 2017. Available online: https://www.congress.gov/bill/115th-congress/senate-bill/680/text (accessed on 15 December 2017).

97. Motor Vehicle Management Act South Korea, 2017; Vol. Act No. 14546. Available online: http://elaw.klri.re.kr/eng_service/lawView.do?hseq=35841&lang=ENG (accessed on 15 December 2017).

98. Son, D.; Sun, H.K. *Self-Driving Cars: New Standard for Data Privacy Internationally and in Korea*; International Bar Association: Seoul, Korea, 2016.

99. Reuters. Germany adopts self-driving vehicles law. *Reuters*, 12 May 2017.

100. Olson, P. Rise of Les Machines: France's Macron Pledges $1.5 Billion to Boost AI. *Forbes*, 29 March 2018. Available online: https://www.forbes.com/sites/parmyolson/2018/03/29/frances-macron-billion-data-sharing-ai/#5cbd84774921 (accessed on 1 April 2018).

101. National Transport Commission Australia. *Regulatory Reforms for Automated Vehicles*; National Transport Commission Australia: Melbourne, VIC, Australia, November 2016.

102. House of Representatives Standing Committee on Industry, I., Science and Resources. *Social Issues Relating to Land-Based Automated Vehicles in Australia*; Parliament of Australia: Canberra, ACT, Australia, 2017.

103. Department for Transport United Kingdom. *The Key Principles of Vehicle Cybersecurity for Connected and Automated Vehicles*; Department for Transport United Kingdom: London, UK, 2017.

104. CCAV. *UK Connected and Autonomous Vehicle Research and Development Projects 2017*; Centre for Connected & Autonomous Vehicles: London, UK, 2017.

105. German Data Protection Watchdog Makes Recommendations for Autonomous and Connected Cars. Avaliable online: https://www.out-law.com/en/articles/2017/june/german-data-protection-watchdog-makes-recommendations-for-autonomous-and-connected-cars/ (accessed on 15 December 2017).

106. Senate Bill 1298, Padilla. Vehicles: Autonomous Vehicles: Safety and Performance Requirements. California, U.S, 2012. Available online: https://leginfo.legislature.ca.gov/faces/billNavClient.xhtml?bill_id=201120120SB1298 (accessed on 22 December 2017).

107. *DMV Releases Draft Requirements for Public Deployment of Autonomous Vehicles*; State of California Department of Motor Vehicles: Sacramento, CA, USA, 2015.

108. *Modified Express Terms Title 13, Division 1, Chapter 1 Article 3.8—Deployment of Autonomous Vehicles—Deployment of Autonomous Vehicles*; State of California Department of Motor Vehicles: Sacramento, CA, USA, 2018.

109. *Regular Session*; House Bill 248; Georgia, U.S, 2017. Available online: http://www.legis.ga.gov/Legislation/en-US/display/20172018/HB/248 (accessed on 22 December 2017).

110. *Regular Session*, 2017–2018 ed.; House Bill 3422; National Conference of State Legislatures: Boston, MA, USA, 2017.

111. Act No. 333 2016. In *SB 0996*; State of Michigan, 2016. Available online: https://www.legislature.mi.gov/documents/2015-2016/publicact/pdf/2016-PA-0333.pdf (accessed on 22 December 2017).

112. Automated Vehicles Act. State of Tennessee, United States; Vol. SB 0151. Available online: http://wapp.capitol.tn.gov/apps/BillInfo/Default.aspx?BillNumber=SB0151 (accessed on 22 December 2017).

113. House Bill 3475. Regular Session. Texas, U.S, 2017. Available online: https://capitol.texas.gov/tlodocs/85R/billtext/pdf/HB03475I.pdf#navpanes=0 (accessed on 22 December 2017)

114. Senate Bill 547. 2017. Available online: http://wapp.capitol.tn.gov/apps/BillInfo/Default.aspx?BillNumber= SB0547&GA=110 (accessed on 22 December 2017).

115. Senate Bill 179. Regular Session. Massachusetts, U.S, 2017. Available online: https://legiscan.com/MA/text/S179/2017 (22 December 2017).

116. House Bill 1829. Regular Session. Massachusetts, U.S, 2017. Available online: https://malegislature.gov/Bills/190/H1829 (accessed on 22 December 2017).

117. PennDOT. *Pennsylvania Autonomous Vehicle Testing Policy: Final Draft Report of the Autonomous Vehicle Policy Task Force*; Pennsylvania Department of Transportation (PennDOT): Harrisburg, PA, USA, 2016.

118. ISO, International Organization for Standardization. *ISO/IEC 27002:2005(e) Information Technology—Security Techniques—Code of Practice for Information Security Management*; International Organization for Standardization: Geneva, Switzerland, 2005.

119. Von Solms, R.; Van Niekerk, J. From information security to cybersecurity. *Comput. Secur.* **2013**, *38*, 97–102.

120. Lee, C. Grabbing the wheel early: Moving forward on cybersecurity and privacy protections for driverless cars. *Fed. Commun. Law J.* **2017**, *69*, 25–52.

121. Parkinson, S.; Ward, P.; Wilson, K.; Miller, J. Cyber threats facing autonomous and connected vehicles: future challenges. *IEEE Trans. Intell. Transp. Syst.* **2017**, *18*, 2898–2915.

122. Yağdereli, E.; Gemci, C.; Aktaş, A.Z. A study on cyber-security of autonomous and unmanned vehicles. *J. Def. Model. Simul.* **2015**, *12*, 369–381.

123. Iacono, M.; Marrone, S. Risk assessment and monitoring in intelligent data-centric systems. In *Security and Resilience in Intelligent Data-Centric Systems and Communication Networks*; Elsevier: New York, NY, USA, 2018; pp. 29–52.

124. Bordonali, C.; Ferraresi, S.; Richter, W. *Shifting Gears in Cybersecurity for Connected Cars*; Mckinsey Company: New York, NY, USA, 2017.

125. NHTSA. *Nhtsa and Vehicle Cybersecurity*; National Highway Traffic Safety Administration: Washington, DC, USA, 2018.

126. U.S. Department of Transportation. *Automated Driving Systems 2.0 a Vision for Safety*; National Highway Traffic Safety Administration: Washington, DC, USA, 2017.

127. Senate Bill 1945. Regular Session. Massachusetts, U.S, 2017. Available online: https://malegislature.gov/Bills/190/SD1195 (accessed on 22 December 2017).

128. Senate Bill 427. Regular Session. Pennsylvania, U.S, 2017. Available online: http://www.legis.state.pa.us/CFDOCS/Legis/PN/Public/btCheck.cfm?txtType=PDF&sessYr=2017&sessInd =0&billBody=S&billTyp=B&billNbr=0427&pn=0396 (accessed on 22 December 2017).

129. Senate Bill 315. 2018. Available online: http://www.legis.ga.gov/legislation/en-US/Display/20172018/SB/315 (accessed on 22 December 2017).

130. Texas Cybersecurity Act. 2017. Available online: https://capitol.texas.gov/tlodocs/85R/billtext/pdf/HB00008F. pdf#navpanes=0 (accessed on 10 January 2018).

131. National Conference of State Legislatures. Cybersecurity Legislation 2017. 2017. Available online: http://www.ncsl.org/research/telecommunications-and-information-technology/cybersecurity-legislation-2017.aspx (accessed on 10 January 2018).

132. Senate Bill 632. Michigan Legislature. Available online: http://www.legislature.mi.gov/(S(ghmwucmvvmso5lij4f1hy1wl))/mileg.aspx?page=GetObject&objectnam e=2017-SB-0632 (accessed on 10 January 2018).

133. The Directive on Security of Network and Information Systems (NIS Directive). Available online: https://ec.europa.eu/digital-single-market/en/network-and-information-security-nis-directive (accessed on 10 January 2018).

134. ENISA. *Cybersecurity and Resilience of Smart Cars*; ENISA: Heraklion, Greece, 2017.

135. KPMG. *Overview of China's Cybersecurity Law*; KPMG Advisory (China) Limited: Huazhen, China, 2017.
136. Yan, S. *China's New Cybersecurity Law Takes Effect Today, and Many Are Confused*; CNBC: Englewood Cliffs, NJ, USA, 2017.
137. Ministry of Home Affairs, Government of Singapore. *Amendments to the Computer Misuse and Cybersecurity Act to Take Effect from 1 June 2017*; 2017. Available online: https://sso.agc.gov.sg/Act/CMCA1993 (accessed on 10 January 2018).
138. European Road Transport Research Advisory Council (ERTRAC). *Automated Driving Roadmap*; European Road Transport Research Advisory Council (ERTRAC) Working Group "Connectivity and Automated Driving": Brussels, Belgium, 2017.
139. Cabinet Office; National Security and Intelligence; HM Treasury; The Rt Hon Philip Hammond MP. *National Cybersecurity Strategy 2016–2021*; HM Government United Kingdom: London, UK, 2016.
140. *Standards for Cybersecurity on New Cars Gets the Green Light*; My News Desk: Stockholm, Sweden; Thatcham Research: Thatcham, UK, 2017.
141. Introduction to the NIS Directive. 2018. Available online: https://www.ncsc.gov.uk/guidance/introduction-nis-directive (accessed on 12 February 2018).
142. Fouquet, H.; Mawad, M. Macron Wants Europe to Forget Facebook Fears and Embrace A.I. Bloomberg. 2018. Available online: https://www.bloomberg.com/news/articles/2018-03-21/macron-wants-europe-to-forget-facebook-fears-and-embrace-a-i (accessed on 3 April 2018).
143. Minter, A. China Could Steer Self-Driving Cars. Bloomberg. 2018. Available online: https://www.bloomberg.com/view/articles/2018-02-04/why-china-could-seize-the-lead-in-self-driving-cars (accessed on 3 April 2018).
144. Ramsey, C.; Wootliff, B. China's Cyber Security Law: The Impossibility of Compliance? *Forbes*, 29 May 2017. Available online: https://www.forbes.com/sites/riskmap/2017/05/29/chinas-cyber-security-law-the-impossibility-of-compliance-2/#42b83ca2201c (accessed on 3 April 2018).